\documentclass[conference]{IEEEtran}
\IEEEoverridecommandlockouts
\usepackage{cite}
\usepackage{url}
\usepackage{amsmath,amssymb,amsfonts}
\usepackage{algorithmic}
\usepackage{graphicx}
\usepackage{textcomp}
\usepackage{xcolor}
\def\BibTeX{{\rm B\kern-.05em{\sc i\kern-.025em b}\kern-.08em
    T\kern-.1667em\lower.7ex\hbox{E}\kern-.125emX}}

\newcommand{\Comment}[1]{}

\begin{document}

\title{Facilitating Change Implementation for Continuous ML-Safety Assurance  \\
\thanks{The first two authors contributed equally to this work. This work was funded by 1) the StMWi Bayern to support the thematic development of Fraunhofer IKS, and 2) Fraunhofer-Siemens Research Project on ML Safety for AGV.}\vspace{-3mm}
}

\author{\IEEEauthorblockN{Chih-Hong Cheng\IEEEauthorrefmark{1},  Nguyen Anh Vu Doan\IEEEauthorrefmark{1}, Balahari Balu\IEEEauthorrefmark{1}, Franziska Schwaiger\IEEEauthorrefmark{1}, Emmanouil Seferis\IEEEauthorrefmark{1}, \\ Simon Burton\IEEEauthorrefmark{1}, 
Yassine Qamsane\IEEEauthorrefmark{2}, 
Ankit Shukla\IEEEauthorrefmark{2}, 
Yinchong Yang\IEEEauthorrefmark{2}, 
Zhiliang Wu\IEEEauthorrefmark{2},
Andreas Hapfelmeier\IEEEauthorrefmark{2},
Ingo Thon\IEEEauthorrefmark{2} }	\IEEEauthorblockA{		\IEEEauthorrefmark{1}Fraunhofer Institute for Cognitive Systems}	\IEEEauthorblockA{		\IEEEauthorrefmark{2}Siemens AG}}

\maketitle

\begin{abstract}

We propose a method for deploying a safety-critical machine-learning component into continuously evolving environments where an increased degree of automation in the engineering process is desired. We associate semantic tags with the safety case argumentation and turn each piece of evidence into a quantitative metric or a logic formula. With proper tool support, the impact can be characterized by a query over the safety argumentation tree to highlight evidence turning invalid. The concept is exemplified using a vision-based emergency braking system of an autonomous guided vehicle for factory automation.

\end{abstract}


\vspace{-3mm}

\section{Problem Statement}

The use of machine learning (ML) in safety-critical applications has moved beyond autonomous driving and entered new domains such as avionics. Although the state-of-the-art paradigm considers the safety engineering of ML to be a continuous engineering process (cf. ISO 21448~\cite{SOTIF} appendix D), one of the core challenges is to accelerate the continuous engineering paradigm. This can be reflected in our use case of integrating ML-based vision for an automated guided vehicle (AGV) within the domain of factory automation. Each factory can have different environmental conditions such as the width of the corridor, and the integrated mobility platforms may have different resource consumption constraints and operational speed. Altogether, when the ML component is safe to use on one shop floor with a particular robot configuration, the transition from one shop floor to another may imply that the established safety argument is no longer valid.

\section{Automating Continuous ML-Safety Assurance}

We present a workflow on how to extend from standard ML-Ops (i.e., set of practices that aims to deploy and maintain ML models in production) with an explicit connection to the assurance case.

\paragraph{Baseline} The following list of information can be considered as a basis required for the ML to be used in a safety-critical application, i.e., the basic snapshot to be demonstrated to the certification authority.

\begin{enumerate}
\item The current ML system (including the pre- and post-processor) being deployed.  

\item The safety case (in the form of Goal-Structuring-Notation - GSN~\cite{kelly2004goal}) associated with the ML on how one argues that the ML engineering process is rigorous and the ML output insufficiency is not a source of creditable harm. 

\item For each evidence listed in the safety case, the associated document or results produced by 3rd party tools demonstrating that the evidence is fulfilled.  

\end{enumerate}

\paragraph{Discovering potential change requests} The first step of the workflow is to know when a change will occur. Newly acquired information such as publicly available incident reports, context evolution (change of factory setup or regulations), or onboard monitoring (e.g., near-collision or out-of-distribution detection) can be used as sources to start the change analysis.

\vspace{1mm}
\paragraph{Automation via formal specification, semantic tags, and evidence evaluation}

When a potential change is uncovered, its impact on the safety case must be assessed. Recall that a GSN maintains a tree-like structure specifying why the deployed ML is safe. Edges are introduced to connect a particular goal to the underlying supporting evidences or decomposed sub-goals. Therefore, understanding the change impact can be reduced to examining the validity of each evidence or assumption located at the leaf of the GSN tree. 

To automate the process, we try to formalize each evidence and assumption using mathematical logic or measurable metrics, such that one can precisely evaluate if a particular piece of evidence, under the change, has turned invalid / false. As an example, the original GSN may contain a claim stating that \emph{the false-positive is only a performance issue rather than a safety issue}. We sharpen the supporting argument using mathematical logic (here with substantial simplification) and reach the following statement:

\vspace{-2mm}
\begin{equation}\label{eq:FPisfine}
\begin{split}
    \forall t: d_{AGV, agent_{rear}}(t) \geq d_{B, agent_{rear}} \\ \wedge \; FP_{ML}(t, img) \wedge detected_{fusion}(t)\\ \rightarrow (\forall t' \in [t, t+ t_{B, AGV}]: d_{AGV,agent_{rear}}(t') > 0)
\end{split}
\end{equation}
\vspace{-2mm}

\begin{table*}[t]
\centering
\caption{Predicates and the associated semantic tags for the translated formula}\label{table:predicate}
\vspace{-2mm}
\begin{tabular}{|l|l|l|}\hline
\textbf{Predicate / Constant} & \textbf{Explaination} & \textbf{Semantic Tags}  \\\hline
$d_{AGV, agent_{rear}}(t)$ & Distance between AGV and the rear agent at time $t$ & distance, AGV, rear-agent  \\\hline
$d_{B, agent_{rear}}$ & Maximum braking distance for the rear agent &  rear-agent, braking distance \\\hline
$FP_{ML}(t, img)$ & With input $img$ at time $t$, the ML has produced a false positive &  false positive, detection \\\hline
$detected_{fusion}(t)$ & Sensor fusion indicates that an object is in the front and close-by & fusion, detection \\\hline
$t_{B, AGV}$ &  Maximum required braking time for the AGV & AGV, braking time \\\hline
\end{tabular}

\vspace{-5mm}

\end{table*}

The explanation of each predicate or constant is shown in Table~\ref{table:predicate}. Intuitively, the formula states an assumption where the rear agent should keep a safe distance, such that when the ML component detects a non-existent object and the AGV triggers a stop, it is impossible for the AGV to be hit from behind.

While the formal specification can assist in understanding the impact over each evidence, the overall activity can also be challenging when one needs to examine each evidence in the complex argumentation tree. We therefore associate each evidence or assumption with a set of \emph{\textbf{semantic tags}} (cf. Table~\ref{table:predicate} for all semantic tags used in Formula~\ref{eq:FPisfine}), store all relevant information using a database, and utilize queries to automatically highlight evidences to be considered.

Finally, by concretizing evidences into mathematical formula, it is also possible to \emph{\textbf{directly evaluate the validity of the formula under change}}. As an example, the braking distance of the AGV is associated with the initial speed of the AGV, which can be formulated using Newton's laws. As safety can be characterized using the travelled distance of the AGV and the agent until the AGV reaches a full stop (similar concept available at responsibility-sensitive safety~\cite{shalev2017formal}), one can derive\footnote{Simple automatic dynamic derivation can be implemented via \texttt{eval()} in python, while complex derivation requires automatic theorem provers.} that increasing the detection frame rate (which is a constant stored in the argumentation) only reduces the travel distance of the AGV and safety is thus not compensated.

\paragraph{Incorporating changes}

To incorporate changes and to again make the ML-system arguably safe, we consider a three stage process with increased difficulty. 

\begin{enumerate}
    \item The simplest case is related to the situation where, by analyzing the safety case, each listed evidence remains valid. One can directly change the parameters stored in the safety case and close the task\footnote{As there can be potential gaps in the argumentation, the impact analysis may be further complemented by re-doing (part of) the tests conducted in the evidence construction process.}.
    
    \item Subsequently, consider if it is possible to perform improvement (e.g., collect more data) such that each evidence listed in the safety case turns valid again. Tools with intelligent test case generation capabilities (e.g.,~\cite{cheng2019nn}) can increase the degree of automation. 
    
    \item Finally, situations exist where it is impossible to incorporate the change by keeping the same assurance case structure\footnote{As an example, when one changes the ML function from object detection to semantic segmentation, as the set of ML output insufficiencies are different the safety argumentation needs to be reworked.}. When the assurance case
structure needs substantial change, human-in-the-loop
modification may be unavoidable.

\end{enumerate}

\vspace{-1mm}
\section{Concept Validation}
\vspace{-1mm}

\begin{figure}[t]
\centering
\includegraphics[width=0.95\columnwidth]{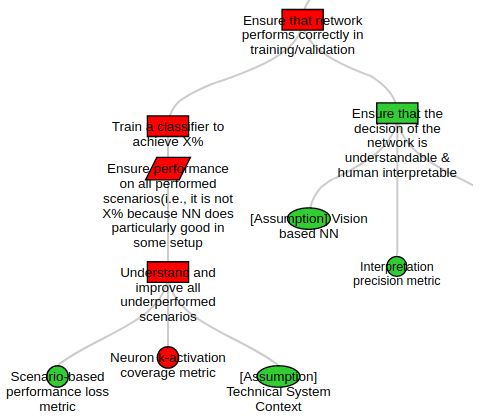}
\vspace{-4mm}
\caption{Validating the concept with an assist of a front-end, where evidence (circle at the bottom left) being impacted is highlighted by the tool in red. }
\label{figure:gsn}
\vspace{-5mm}
\end{figure}

We have built a research prototype 
with the help of multiple open-source tools. We used D3.js 
for building a front-end visualizing the complete GSN tree (Figure~\ref{figure:gsn} shows an example from~\cite{cheng2019nn}  where one piece of evidence has turned invalid), where the tree and all evidences are dynamically linked to a NoSQL database.
Apart from formally specified formulae, we also connected the evidence with the ML testing tool \emph{nn-dependability-kit}~\cite{cheng2019nn} which can provide various types of metrics such as data completeness.  

The database allows easy query for semantic tags to identify nodes in the argumentation tree that a change may impact. Additionally, it allows to flexibly update the status (true or false) of an evidence, either automatically (via the automatic evaluation of an embedded formula), with an external connection to the database (3rd party tool used to demonstrate an evidence is fulfilled), or manually when required. 

\section{Concluding Remarks}

Our proposed methodology aims at the efficient iteration of the ML-Ops where safety argumentation is an integral part of the engineering. The methodology is particularly suitable in the initial prototyping phase, where safety should be part of the engineering process, but the system architecture can continuously evolve. Realizing the concept into tools, the result also facilitates communication between ML engineers and safety analysts working on seemly unrelated domains. It also leads to various improvement potentials, such as using natural language processing to reduce human efforts further.  


\bibliographystyle{abbrv}

\begin{thebibliography}{1}
	
	\bibitem{SOTIF}
	{ISO } 21448:2022 road vehicles - safety of the intended functionality.
	\newblock \url{https://www.iso.org/standard/77490.html}, 2022.
	
	\bibitem{cheng2019nn}
	C.-H. Cheng, C.-H. Huang, and G.~N{\"u}hrenberg.
	\newblock nn-dependability-kit: Engineering neural networks for safety-critical
	autonomous driving systems.
	\newblock In {\em ICCAD}, pages 1--6. IEEE, 2019.
	
	\bibitem{kelly2004goal}
	T.~Kelly and R.~Weaver.
	\newblock The goal structuring notation--a safety argument notation.
	\newblock In {\em DSN Workshop on Assurance Cases}, page~6. Citeseer, 2004.
	
	\bibitem{shalev2017formal}
	S.~Shalev-Shwartz, S.~Shammah, and A.~Shashua.
	\newblock On a formal model of safe and scalable self-driving cars.
	\newblock {\em arXiv preprint 1708.06374}, 2017.
	
\end{thebibliography}

\end{document}